\documentclass{sig-alternate-10pt}

\usepackage{graphicx}   % extended graphics package
\usepackage{amsmath}    % math package
\usepackage{cases}
\usepackage{hhline}     % extended styles for tables
\usepackage{cite}
\usepackage{times}
\usepackage[font=footnotesize,labelfont=bf]{caption}

\makeatletter
\def\ps@headings{%
\def\@oddhead{\mbox{}\scriptsize\rightmark \hfil \thepage}%
\def\@evenhead{\scriptsize\thepage \hfil \leftmark\mbox{}}%
\def\@oddfoot{}%
\def\@evenfoot{}}

\begin{document}
\title{Modeling and Improving the Energy Performance of GPS Receivers for Mobile Applications
\titlenote{This work was supported in part by the SIAT Innovation Program for Excellent Young Researchers (201307).}}

\numberofauthors{1}
\author{
\alignauthor Kongyang~Chen and Guang~Tan\\
\affaddr{SIAT, Chinese Academy of Sciences, China.}\\
\affaddr{\{ky.chen, guang.tan\}@siat.ac.cn}}

\maketitle

\begin{abstract}
Integrated GPS receivers have become a basic module in today's
mobile devices. While serving as the cornerstone for location based
services, GPS modules have a serious battery drain problem due to
high computation load. This paper aims to reveal the impact of key
software parameters on hardware energy consumption, by establishing
an energy model for a standard GPS receiver architecture as found in
both academic and industrial designs. In particular, our
measurements show that the receiver's energy consumption is in large
part linear with the number of tracked satellites. This leads to a
design of selective tracking algorithm that provides similar
positioning accuracy (around 12m) with a subset of selected
satellites, which translates to an energy saving of 20.9-23.1\% on
the Namuru board.
\end{abstract}

\section{Introduction}\label{sec:introduction}

The Global Positioning System (GPS) is one of the key technologies
that have shaped today's mobile Internet. As a cornerstone for
location based services, integrated GPS receivers have become a
standard module in mobile devices. A main problem with GPS receivers
is that they are very power
hungry~\cite{entracked,rate-adaptive,mobile-loc,sensloc,smartloc,sfft,cloud-offload,leap}
-- a typical GPS module consumes energy in the range of 143 to 166
mW~\cite{power-smartphone} in the continuous navigation model, which
would deplete a mobile phone's battery in merely six hours.

Various techniques have been proposed to address this problem. The
hybrid location sensing
technique~\cite{radio-sensys,enloc,entracked,rate-adaptive,mobile-loc,location-sensing,sensloc,smartloc}
uses alternative positioning methods such as cell tower
triangulation, WiFi, radio, or accelerometers to help the terminal
reduce GPS sampling frequency. The drawback is that these helper
techniques can greatly increase positioning errors, sometimes up to
hundreds of meters~\cite{rate-adaptive}. The second technique uses a
sparse Fast Fourier Transform (FFT) method~\cite{sfft} to reduce the
amount of computation in the receiver software, in order to lower
energy consumption. However, for a GPS receiver, FFT is only
required during the satellite acquisition phase, whose amortized
load is quite low during continuous sampling, thus the overall
energy saving is insignificant. In~\cite{cloud-offload}, high
complexity computational workloads are offloaded from the receiver
to a cloud server to reduce the energy consumption. This approach
prevents the solution from being useful in real-time navigation
applications.

In this paper we explore a new approach to improving the energy
efficiency of a GPS receiver. Different from the hybrid location
sensing and cloud offloading approaches, we do not assume external
hardware (e.g., inertial sensors), but focus on the internal
structure and characteristics of the receiver, aiming to offer a
transparent energy saving solution for upper layer applications. It
is also different from the sparse FFT approach in that we do not
limit our attention of a specific (and small) part of the
computation task, but consider the whole process of signal
processing and position calculation. The intuition that motivated
our study is that there exists significant redundancy of satellite
information among successive cycles of position calculation on a
mobile device. By using this redundancy, some computation may be
saved and thus the energy consumption reduced.

To evaluate the impact of computation efficiency on energy
efficiency, we need an energy model to relate algorithm performance
with hardware energy consumption. To the best of our knowledge,
there has been no published work that addresses this problem,
probably due to two challenges. First, modern GPS receivers on
mobile devices are quite complex in structure, comprising an array
of hardware units including antenna, radio front end (RF), digital
signal processor, main processor, as well as memory in different
forms. The overall energy performance depends on energy expenses of
these individual units, whose characteristics and interconnections
vary greatly across brands and models. It is therefore difficult to
obtain an accurate yet general energy model. Second, the computation
of GPS involves multiple complicated procedures executed in an
interleaving fashion; it requires a thorough test and analysis to
disentangle key components and parameters from the collective
performance of the whole system.

\begin{figure*} [t]%[tb]
\begin{minipage}{0.65\linewidth}
\centering
\includegraphics[width=0.9\hsize]{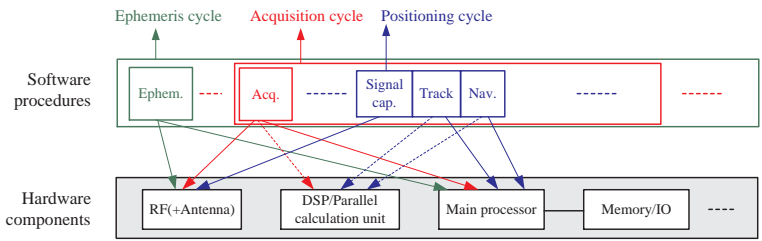}
\caption{Architecture of a typical GPS receiver: main software
  procedures and associated hardware
  components. Dashed arrow lines mean possible association, depending on
  implementation.}\label{fig:architecure}
\end{minipage}
\hspace{10pt}
\begin{minipage}{0.30\linewidth}
%\begin{table}
\centering \footnotesize
\begin{tabular}{lp{0.8cm}p{0.8cm}p{0.8cm}}\hline
   Module/State       &    Power     &   Duration    & Period\\\hline

    RF & $P_{r}$     &     $t_{r}$     &     $T$\\\hline

    Acquisition  & $P_a$ & $t_a$     &     $T_a$\\\hline

    Track        & $P_t$     & $t_t$     & $T$\\\hline

    Ephemeris    & $P_e$     & $t_e$ & $T_e$\\\hline

    Navigation   & $P_n$     & $t_n$     & $T$\\\hline

    System Idle  &    $P_i$     &     $t_i$     &     $T$\\\hline
\end{tabular}
\caption{Energy parameters of main software procedures and associated power levels.}
\label{table:energy-model}
%\end{table}
\end{minipage}
\end{figure*}

We present an energy model that addresses the above challenges based
on measurement with the Namuru V2 GPS receiver~\cite{namuru}. The
model captures the architecture of a typical GPS receiver, as found
in research-oriented~\cite{namuru} and industrial
designs~\cite{SiRFstarIV,sirf-patent,ublox}, while ignoring platform
specific details and optimizations. The goal is not to predict
absolute energy consumption of a general GPS module, but to shed
light on the relationship between energy consumption and major
software strategies. For the case of Namuru, a most notable finding
is that the receiver's energy consumption is roughly linear with the
number of satellites to be tracked. Based on this finding, we
propose a selective satellite tracking algorithm that minimally
synchronizes with visible satellites, by taking advantage of the
short term stability of satellite signal quality. Compared with the
traditional full tracking algorithm, we obtain significant energy
savings with negligible sacrifice on positioning accuracy. In
summary, this paper makes two contributions:

\begin{itemize}
\item An energy model for GPS receivers showing the major
energy consumers, and revealing the relationship between energy
consumption and key software parameters. The model allows one to
focus on the optimization of certain parts of GPS receiver software,
which can be conveniently translated to energy gains. Being the
first of its kind, the model provides a basis for our future
investigation of a GPS receiver's energy performance.

\item A tracking algorithm that opportunistically avoids unnecessary
satellite tracking for positioning. Real traces in two cities show
that our new algorithm can save 20.9-23.1\% energy consumption while
retaining similar positioning accuracy.
\end{itemize}

\section{GPS and GPS Receivers}\label{sec:principle}
The GPS navigation system is constituted of three components,
satellites constellation, ground stations, and user receivers.
The satellites constellation contains 32 satellites orbiting
the Earth every 12 hours~\cite{borre,kaplan}. The ground stations
keep tracking the satellites' health and trajectory configuration
including the almanac and the ephemeris, which indicate the satellite's
status and precise location. All the satellites are precisely
synchronized to atomic clocks within a few nanoseconds. Each
satellite continuously broadcasts its time and trajectory message
with CDMA signals at L1=1.575GHz (or L2=1.227GHz). GPS receivers
capture raw GPS signals, decode the carrier/code information,
and calculate their three-dimensional locations with a least
square method.

Figure~\ref{fig:architecure} shows the architecture of a typical GPS
receiver design. From the perspective of software, the receiver
mainly consists of five software procedures that are executed in
cycles of different periods. In theory, the ephemeris data included
in the satellite broadcast is only valid for 30 minutes, so the
ephemeris data needs to be collected and decoded every 30 minutes.
The acquisition procedure is executed every few minutes to extract
the information of visible satellites. The positioning cycle refers
to the time interval between two position updates to the
application, and determines a system parameter known as {\em update
rate}. Modern GPS modules often provide an update between 1 and 10
Hz. During each positioning cycle, the receiver calculates for a
position, after which the receiver may enter a low-power sleep or
idle state until the start of the next positioning cycle. Each
positioning cycle involves three software procedures: signal capture
and processing, (satellite) track, and navigation.

Each software procedure involves a number of hardware components.
For example, the acquisition procedure is highly computation
intensive and often requires dedicated hardware in addition to its
use of the RF and the main processor. The main processor is in
charge of task scheduling and general processing logic, so it is needed
in all procedures. The specific hardware composition varies greatly
across receiver manufacturers. For example, the SiRFstarIV chipset
uses an ARM7 as the main processor and a DSP for faster signal
processing~\cite{SiRFstarIV}, the ublox LEA-6 module~\cite{ublox}
uses a dedicated hardware engine for massive parallel searches, while
the Namuru receiver defines its CPU and parallel calculation unit
with an Altera Cyclone 2C50 FPGA~\cite{namuru}.

\section{A Generic Energy Model}\label{sec:energy}

In this section we establish a generic energy model for a standard
GPS receiver, assuming the architecture in
Figure~\ref{fig:architecure}. Recall that we have identified five
main procedures that dominate the energy consumption of a receiver:
signal capturing and processing (i.e., the RF), acquisition, track,
ephemeris extraction, and navigation. Some procedures span multiple
positioning cycles (or simple cycles, when no confusion occurs),
while others are on a per-cycle basis, all subject to a scheduler in
the main processor. Given the typical single core configuration of
the main processor, we can approximate the energy consumption of a
procedure executed in scheduled time slots with its energy
consumption during a complete and continuous run. A list of energy
related variables is given in Figure~\ref{table:energy-model}. %Table~\ref{table:energy-model}.

Let $f$ be the update rate of the receiver, then the positioning
cycle is $T=1/f$ seconds. The software procedures have the following
energy characteristics:

\begin{itemize}

\item In each cycle, the RF captures raw GPS signal for a period of $t_r$
time, with a power level $P_r$. Normally, $1 \sim 2$ms raw data
suffices to produce a position~\cite{cloud-offload}, so
$t_r=0.002$s.

\item In the acquisition procedure, the receiver samples GPS signal
for a period of $T_r$ time, and determines which satellites are in
view by correlating the signal with a predefined C/A code, with an
acquisition time $t_a$ and power $P_a$. For faster processing, the
procedure may involve parallel computing units. In theory, the
acquisition needs to be done only once for every period of
ephemeris. However, in practice, the receiver may lose its lock with
some of the satellites, so the actual operation of acquisition
should be more frequent. In our experiments, an acquisition period
of one minute turns out to work well, so we set $T_a=60$s.

\item In the track procedure, the main processor calculates the precise Doppler
frequency and code phase for each positioning cycle, with a track
time $t_t$ and power $P_t$.

\item In the ephemeris extraction procedure, the receiver has the RF continuously sample
GPS signal for at least $t_{re}=36s$, and then the main processor calculates
the status, position, and orbits of visible satellites, with a
running time $t_e$ and power $P_e$, for every $T_e = 30$ minutes.
The calculation includes an acquisition and a track procedures.

\item In the navigation procedure, the main processor calculates the receiver's position by a least
square method, with a running time $t_n$ and power $P_n$.

\item The receiver enters an idle state and stays for $t_i$ time after obtaining a location fix in each cycle, with a power level $P_i$.
\end{itemize}

The total energy consumption of a GPS receiver in a time unit is:
\begin{equation}
\begin{aligned}
%&P = P_i \frac{t_i}{T} + P_r \frac{t_r}{T}
%   + P_a \frac{t_a}{T_a} + P_t \frac{t_t}{T} \\
%&\quad
%   + P_e \frac{t_e}{T_e} + P_n \frac{t_n}{T}.
&P = P_i \frac{t_i}{T} + P_r \frac{t_r}{T}
   + \frac{P_a t_a + P_r t_r}{T_a} + P_t \frac{t_t}{T} \\
&\quad
   + \frac{P_e t_e + P_r t_{re}}{T_e} + P_n \frac{t_n}{T}.
\end{aligned}
\end{equation}

Transforming the power to the product of voltage and current gives:
\begin{equation}\label{eq:basic-model}
%P = U_r I_r \frac{t_r}{T} +
%    U_s( I_a \frac{t_a}{T_a} + I_t \frac{t_t}{T}
%       + I_e \frac{t_e}{T_e} + I_n \frac{t_n}{T} ),
%
%P = U_r I_r (\frac{t_r}{T} + \frac{t_r}{T_a} + \frac{t_{re}}{T_e})+
%    U_s( I_a \frac{t_a}{T_a} + I_t \frac{t_t}{T}
%       + I_e \frac{t_e}{T_e} + I_n \frac{t_n}{T} ),
%
\begin{aligned}
& P = U_r I_r \frac{t_r}{T} +
     \frac{U_s I_a t_a + U_r I_r t_r}{T_a} + U_s I_t \frac{t_t}{T} \\
&\quad\quad
    + \frac{U_s I_e t_e + U_r I_r t_{re}}{T_e} + U_s I_n \frac{t_n}{T},
\end{aligned}
\end{equation}
\noindent where $U_r$ and $I_r$ are the voltage and current of the
RF module, $U_s$ the voltage of the main processor, and $I_a$,
$I_t$, $I_e$, $I_n$ are the currents of the FPGA for the
acquisition, track, ephemeris extraction and navigation procedures,
respectively. These currents are different because the procedures
may involve the main processor, parallel computing unit, and other
memory system in different ways.

The system idle state is a deep power saving mode with $2 \sim 3$
lower orders of magnitude than other modules, so it can be omitted.
Plugging $T=1/f$ and the settings $t_r = 0.002$, $T_a=60$, $t_{re}=60$, and
$T_e=1800$ into Eq.~\ref{eq:basic-model}, we have
\begin{equation}\label{eqn:energyModel}
%P = 0.002U_r I_r f+
%    U_s( I_a \frac{t_a}{60} + I_t t_t f
%       + I_e \frac{t_e}{1800} + I_n t_n f).
%
%P = U_r I_r (0.002 f + 0.02)+
%    U_s( I_a \frac{t_a}{60} + I_t t_t f
%       + I_e \frac{t_e}{1800} + I_n t_n f).
%
\begin{aligned}
& P = 0.002 U_r I_r f + \frac{U_s I_a t_a + 0.002 U_r I_r}{60}
   + U_s I_t t_t f\\
&\quad\quad
   + \frac{U_s I_e t_e + 36 U_r I_r}{1800} + U_s I_n t_n f.
\end{aligned}
\end{equation}

\section{The Case for Namuru Receiver}\label{sec:model_namuru}

\begin{figure} [t]%[tb]
\centering
\includegraphics[width=0.9\hsize]{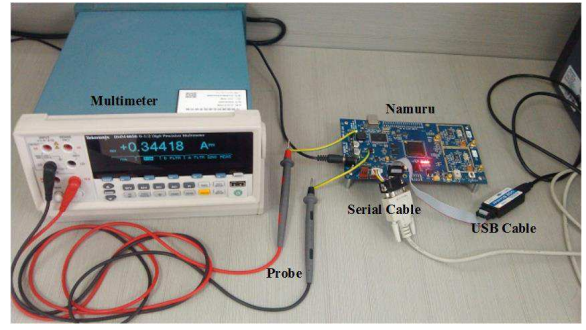}
\caption{Energy parameter measurement system.}
\label{fig:meas-meter}
\end{figure}

We study the energy model using the Namuru V2 receiver~\cite{namuru}
as a concrete example. The goal is to determine the impact of key
software parameters on energy efficiency. Two such parameters are
considered: number of satellites to be tracked, denoted by $N$, and
length of raw data (in milliseconds) to be sampled in each second,
denoted by $L$. The more satellites to be synchronized, the higher
positioning accuracy we get. The length of GPS data sampled per second
determines the receiver's maximum update rate which affects the
resolution of position, since every position fix requires a certain
length of raw data. In our setting, we use 2ms data for a position
fix, therefore we have $L=2f$, in number.

The Namuru board is based on the Altera Cyclone 2C50 FPGA and
contains a RF, RAMs/flashes, as well as IOs. We concentrate on core
components such as the RF and FPGA. The measurement system is shown
in Figure~\ref{fig:meas-meter}. The multimeter DMM4050 is connected
to the Namuru board with its two probes, in order to record the
realtime energy consumption of each module. To measure the voltage, the
multimeter is connected to the RF or FPGA in parallel. To measure the
current, a cascaded 0R resistance in the Namuru board is replaced by the
multimeter probes.

\begin{figure}[t]%[tb]
    \centering
    \includegraphics[width=0.5\linewidth]{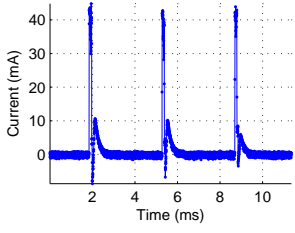}
    \caption{Current of the FPGA when running the track program for $L=2$ (after subtracting
    a constant baseline current).}
    \label{fig:current}
\end{figure}

To study the impact of individual software procedures on energy
consumption, we reorganize the source code of the Namuru
project, and create independent test units with configurable inputs
of $N$ and $L$. These test units are created in the NIOS II IDE,
downloaded to the FPGA, and then measured by the multimeter.
Figure~\ref{fig:current} (a) shows an example of the current
measurement for three runs of the track procedure separated by
intervals, after subtracting a constant baseline current of about
350 mA.

\begin{figure} [t]%[tb]
\begin{minipage}{0.46\linewidth}
\centering
\includegraphics[width=1\hsize]{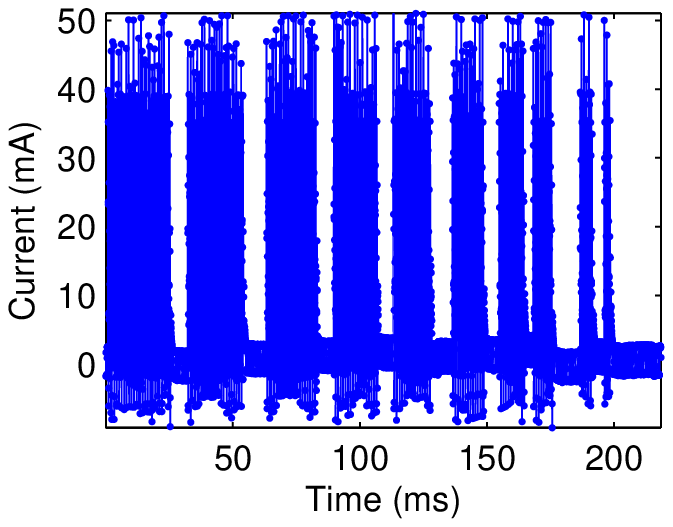}
\caption{Current of FPGA running the track procedure for
$N=10,9,\ldots,1$.} \label{fig:track}
\end{minipage}\hspace{5pt}
\begin{minipage}{0.46\linewidth}
\centering
\includegraphics[width=1\hsize]{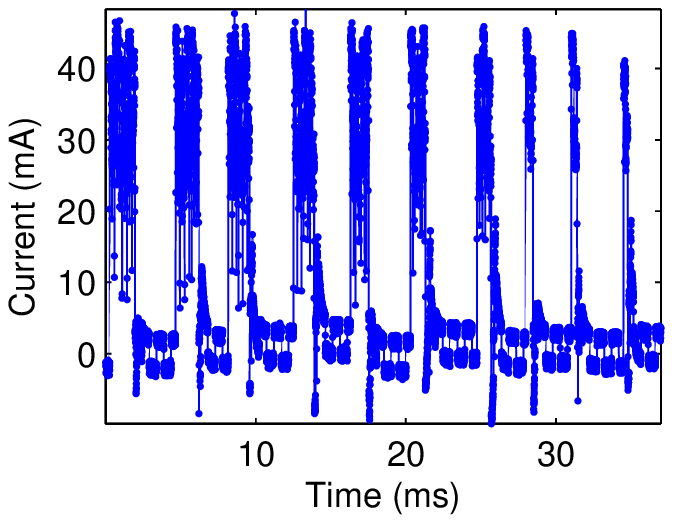}
\caption{Current of FPGA running the navigation procedure for
$N=10,9,\ldots,1$.} \label{fig:navigation}
\end{minipage}
\end{figure}

\begin{figure}[t]%[tb]
    \centering
    {\footnotesize
    %\quad
    \shortstack{
            \includegraphics[width=0.49\linewidth]{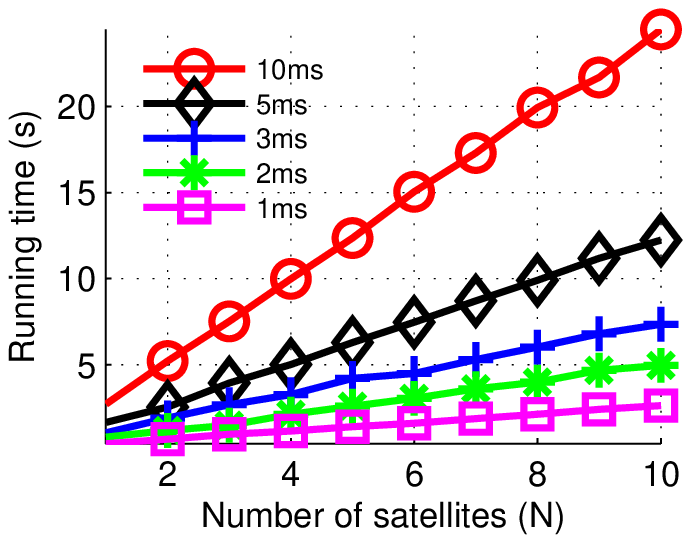}\\
            {(a) Running time}
    }
    %\quad
    \shortstack{
            \includegraphics[width=0.49\linewidth]{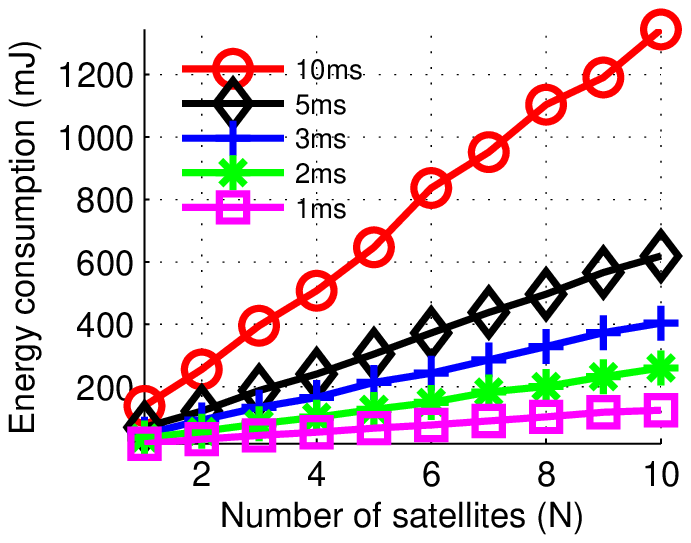}\\
            {(b) Energy consumption}
    }
    }
    \caption{Running time and energy consumption of the track procedure vs. number of tracked satellites.}
    \label{fig:track-results}
\end{figure}

\begin{figure}[t]%[tb]
    \centering
    {\footnotesize
    %\quad
    \shortstack{
            \includegraphics[width=0.49\linewidth]{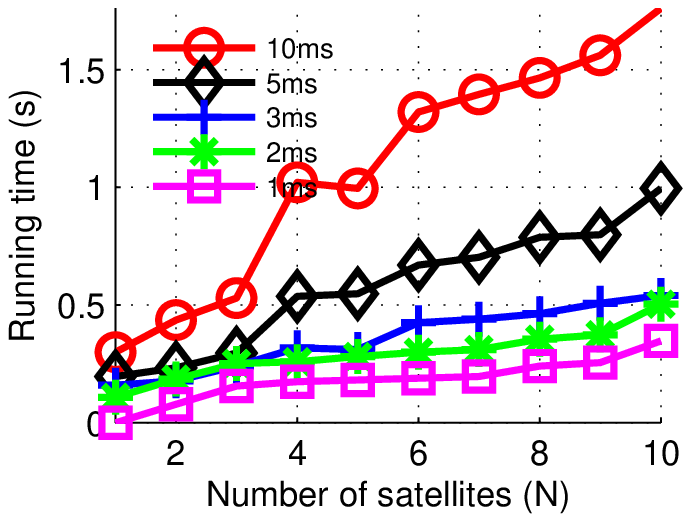}\\
            {(a) Running time}
    }
    %\quad
    \shortstack{
            \includegraphics[width=0.49\linewidth]{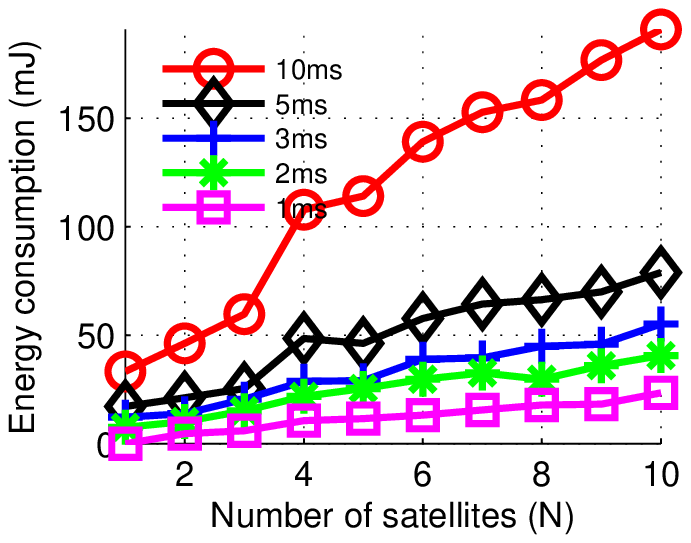}\\
            {(b) Energy consumption}
    }
    }
    \caption{Running time and energy consumption of the navigation procedure vs. number of tracked satellites. }
    \label{fig:navigation-results}
\end{figure}

\begin{figure}[t]%[tb]
    \centering
    \includegraphics[width=0.6\linewidth]{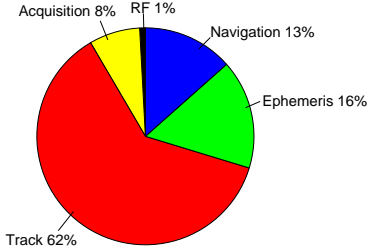}\vspace{-10pt}
    \caption{Average energy consumptions of different procedures.}
    \label{fig:power}
\end{figure}

\textbf{RF.} After removing the 0R resistance, the multimeter
acquires the voltage $U_r=5$V and current $I_r=64$mA. Thus $P_r =
0.002 U_r I_r f = 0.64f mW $. (There are two identical RF circuits
and an additional frequency up-converter circuit on the Namuru board; we
consider only the GPS L1 circuit.)

\textbf{Acquisition.} The acquisition procedure needs to search 30+
Doppler frequency bins and 8000+ code phases for each satellite,
which is a very computation intensive operation. We disable the RF,
import GPS signal traces manually, and run the acquisition
test unit from a cold start. We obtain  the voltage $U_s=3.3$V, the current
$I_a=130$mA, and the running time $t_a=1.2$s. Considering the RF
sampling procedure before each acquisition, the total energy
consumption is $P_a = (U_s I_a t_a + 0.002 U_r I_r)/60 = 8.59$ mW.

\textbf{Track.} A GPS receiver calculates the precise Doppler
frequency and code phase in each positioning cycle. We measure the
track current under different $N$ and $L$. Figure~\ref{fig:track}
shows the raw measurement of current under continuous tracking, for
$L=10$ms and $N=10,9,\ldots,1$. In this figure, each continuous blue
region represents a specific $N$. The current, multiplied by
voltage, integrated over the duration of each such region measures
the energy consumption of a procedure.
Figure~\ref{fig:track-results} shows how running time and energy
consumption depend on $N$ and $L$. As can be seen, the energy
consumption is roughly linear with $N$ and $L$, fitted with $P_t =
11.88 + 7.26 \cdot N \cdot L$ mW.

\textbf{Ephemeris.} We obtain $I_e=131$mA and $t_e=50$s in the cold
start. with the addition of RF sampling, the whole energy
consumption is $P_e = (U_s I_e t_e + 36 U_r I_r)/1800 = 18.4$ mW.

\textbf{Navigation.} Similar to the track procedure, we measure the
energy parameters under different $N$ and $L$.
Figure~\ref{fig:navigation-results} suggests that the energy
consumption is roughly linear with $N$ and $L$, fitted with
$P_n = 2.0 + 1.65 \cdot N \cdot L$ mW.

According to Eq.~\ref{eqn:energyModel}, the energy consumption for
each positioning cycle are $P_r=0.64f$, $P_a=8.59$, $P_t=11.88+7.26
NL$, $P_e=18.4$, $P_n=2.0+1.65NL$, in mW. So, the amortized energy
consumption per second is as follows:

\begin{equation}\label{eq:final-model}
%P = 42.06 + 8.91\cdot N\cdot L = 42.06+17.82\cdot N \cdot f.
\begin{aligned}
&P = 40.87 + 0.64\cdot f + 8.91\cdot N\cdot L \\
&\quad
  = 40.87 + 0.64\cdot f +17.82\cdot N\cdot f.
\end{aligned}
\end{equation}

Figure~\ref{fig:power} presents a breakdown of the different
procedures's energy consumption on Namuru with $N=8$ satellites tracked and $f=1$Hz
location update rate. It can be seen that the dominant energy
consumer is the track procedure, expending up to 62\% of the total
energy. This is partly because this procedure is in the innermost
loop of the processing flow (Figure~\ref{fig:architecure}), and
partly because it involves intensive computation itself.

\section{Energy Efficient Tracking}\label{sec:positioning}
The model in eq.~\ref{eq:final-model} suggests that if one
manage to reduce the number of tracked satellites or update rate,
then the energy consumption can be reduced considerably. The problem
is how to retain the positioning accuracy at the same time. We make
an initial investigation in this section, considering the parameter
$N$, with $f=1$.

\subsection{Selective Tracking}

Traditional track algorithms attempt to track all visible satellites
as indicated by the ephemeris information, and then use all or part
of the satellites to calculate positions. Our approach is to track
only a subset of the visible satellites that are just enough to
produce equally accurate positions. The observation behind is that
the signal quality of satellites remains stable for at least a short
period of time (e.g., minutes), and that the contributions of
available satellites to position quality are nonuniform. Thus, we
can perform full tracking only sparingly, and selectively track a
subset of satellites whose collective quality is close to that of
the full set.

\textbf{Satellite weight.} The geometric dilution of precision
(GDOP) is a scaling factor to show the pseudorange error between the
receiver and available satellites, which can be determined as
follows:
\begin{equation}
\begin{aligned}
& \mbox{GDOP}= \sqrt{\mbox{trace}(A^T A)^{-1} },\\
& A = \begin{bmatrix}
u_1     &  v_1     &  w_1     & 1\\
u_2     &  v_2     &  w_2     & 1\\
\vdots  &  \vdots  &  \vdots  & 1\\
u_r     &  v_r     &  w_r     & 1\\
\end{bmatrix},
\end{aligned}
\end{equation}

\noindent where $A$ is a geometry matrix between the receiver and
available satellites, ($u_i, v_i, w_i)$ is the normalized direction
vector between the receiver and the $i^{th}$ satellite, and $r$ is
the number of available satellites.

Suppose $W$ is a weight matrix to indicate each satellite's
contribution to minimize the GDOP, which equals to minimize the
convariance matrix $(A^T W A)^{-1}$~\cite{gdop-select}.

With $M=A(A^T W A)^{-1}$, it can be proved that
\begin{equation}
\begin{split}
\mbox{trace}(A^T W A)^{-1}  = \mbox{trace}(W M M^T) \\
 = \sum_{i=1}^r \sum_{j=1}^4 w_{ii} m_{ij}^2
  \triangleq f(W).
\end{split}
\end{equation}

Then, differentiate $f(W)$ to minimize $f(W)$ as follows:
\begin{equation}
  \frac{\partial f(W_{kk})}{\partial w_{kk}} = \sum_{j=1}^4 m_{kj}^2 = 0,
\end{equation}
\noindent which can be solved with an iteration method to derive $W_{kk}$.

\textbf{Satellite selection.} In selective tracking, satellites are
selected as follows. Suppose the current satellite set $S$ has a
total GDOP $G$. First, calculate each satellite's contribution to
the positioning accuracy with the above satellite weight algorithm.
Then, choose a satellites subset $S_w$ with three largest weights,
and obtain its GDOP, denoted by $G_w$. This subset is considered
qualified if $ \left| G-G_w\right| / G < 5\% $. Otherwise, add the
largest weight satellite in subset $S-S_w$ to $S_w$, until $ \left|
G-G_w\right| / G < 5\% $. Note that traditional positioning
algorithm requires at least four satellites to determine the
receiver's location. With the historical receiver's location, the
altitude is known and so three satellites are sufficient for
positioning.

We have found that the relative GDOPs of satellites usually remain
stable during intervals of minutes, though their absolute values
vary over time. Thus, the satellite selection algorithm (following a
full tracking operation) only needs to be performed every few
minutes. In our setting, it is executed once a minute. The
computation involves simple operations on small-sized matrices
($r\times 4$ in our case), a small number of iterations for
minimizing $f(W)$ (normally 3, with dynamic adjustment of search
step size), and a greedy selection algorithm, so the per-second
overhead, in terms of both computation load and energy consumption,
is negligible.

\subsection{Evaluation}

\begin{figure}[t]%[tb]
    \centering
    {\footnotesize
    %\quad
    \shortstack{
            \includegraphics[width=0.32\linewidth]{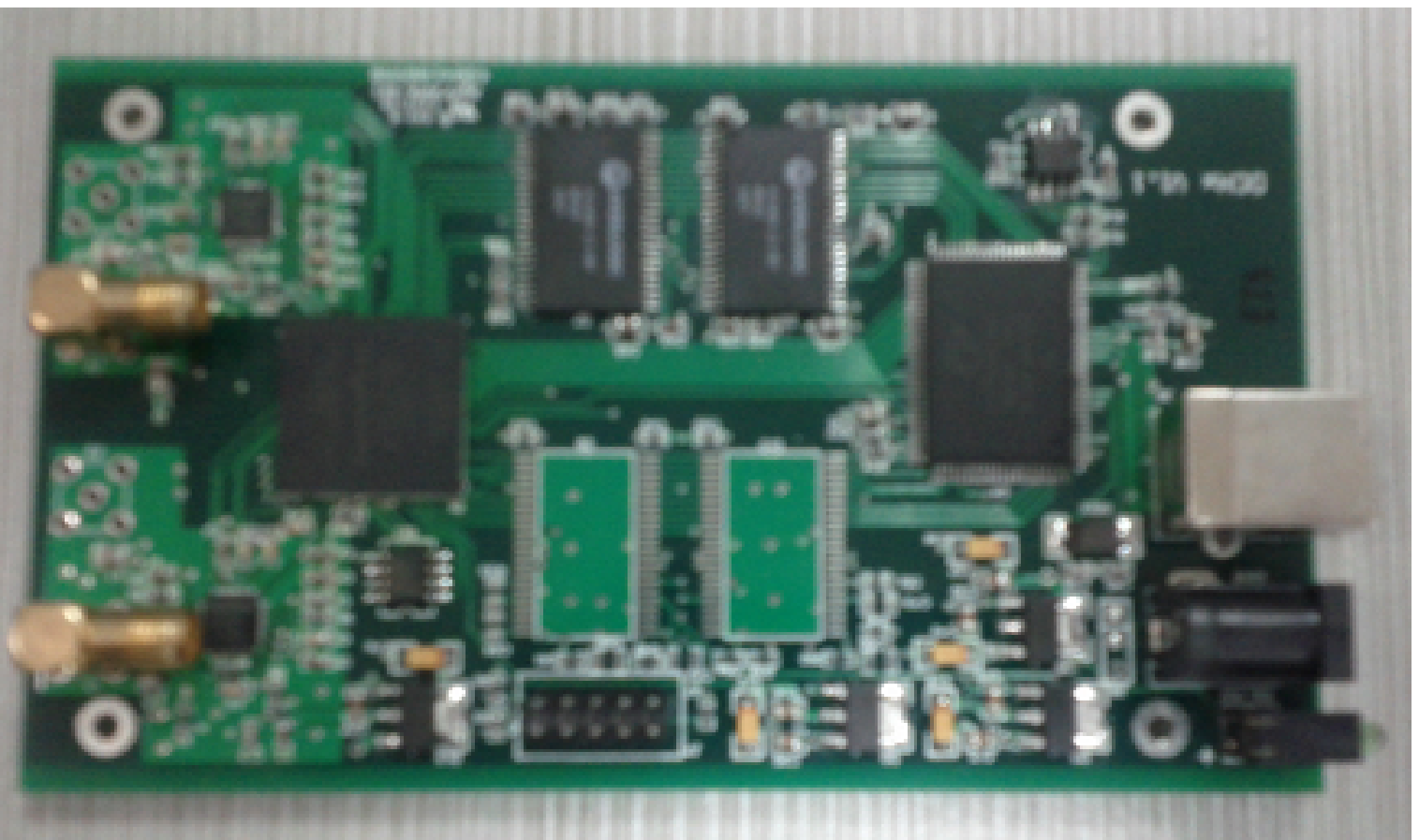}\\
            {(a)}
    }
    %\quad
    \shortstack{
            \includegraphics[width=0.32\linewidth]{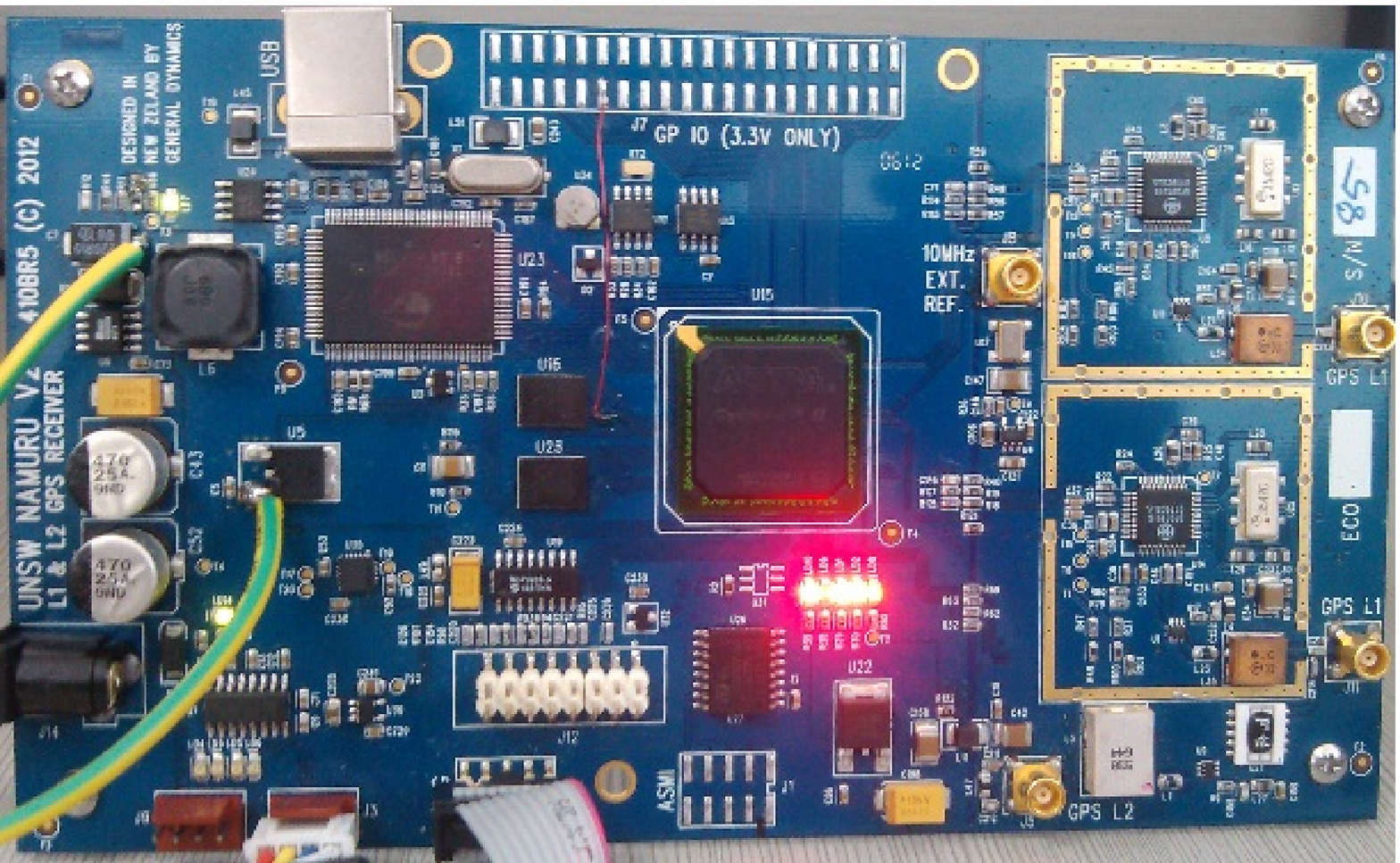}\\
            {(b)}
    }
    %\quad
    \shortstack{
            \includegraphics[width=0.32\linewidth]{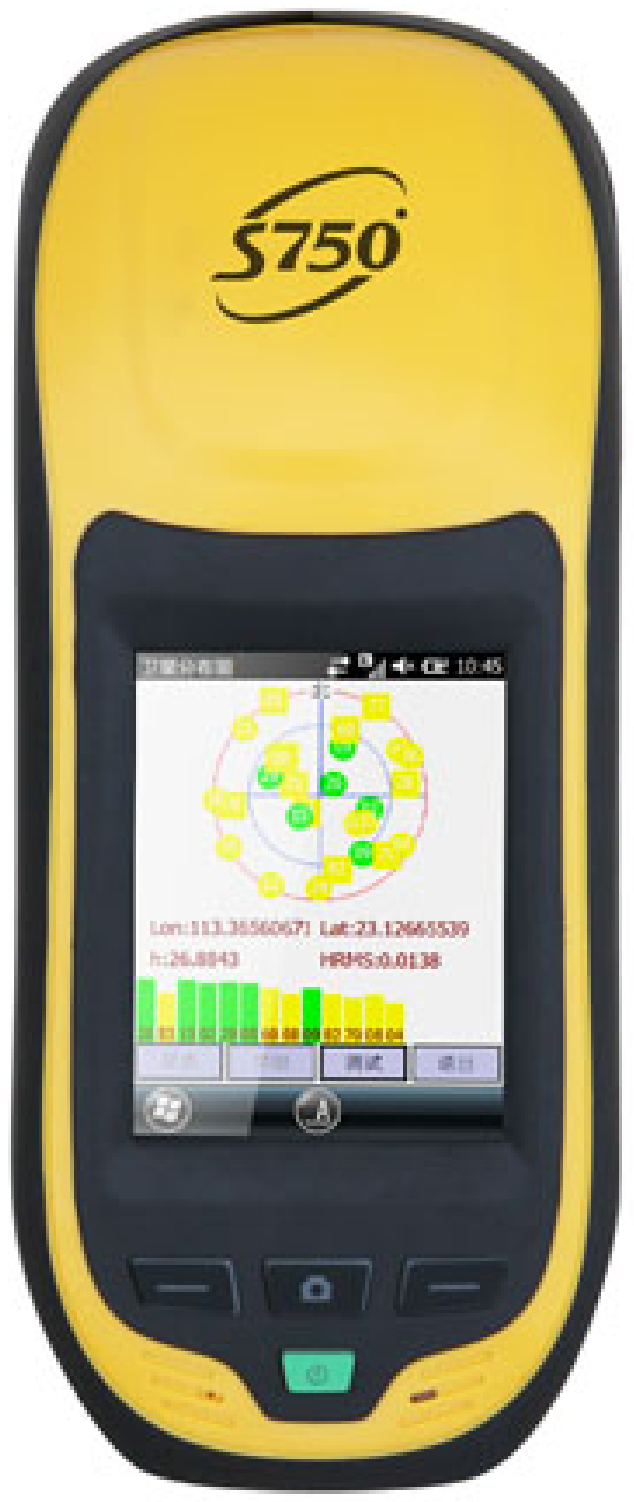}\\
            {(c)}
    }
    }
    \caption{GPS samplers and a professional positioning device for obtaining ground truth. (a) HG-SOFTGPS02 sampler. (b) Namuru. (c) Handheld GIS data collector S750.}
    \label{fig:gps-sampler}
\end{figure}

\textbf{Experimental setup.} We evaluate our algorithm using real
mobile data traces. Two GPS samplers were used,
HG-SOFTGPS02~\cite{HG-SOFTGPS} and Namuru~\cite{namuru}, shown in
Figures~\ref{fig:gps-sampler}(a) and (b). These GPS samplers collect
2bit data, with a sampling frequency 16.368 MHz and an intermediate
frequency 4.092 MHz.

While sampling the data on the vehicle, we used a professional
handheld GIS data collector S750 to obtain the ground truth of
positions~\cite{south-instrument}. As shown in
Figure~\ref{fig:gps-sampler}(c), the collector has a professional
GPS module with post-processed kinematic mode and CORS network
access authority. It provides an update rate of 1 Hz with sub-meter
accuracy. We collected two traces. The first trace is about 4.8 km
long, obtained on a highway with 60 km/h velocity. The full tracking
method finds 6-8 effective satellites on this road, and generates
11.8m location accuracy. The other traces was gathered from a
different city which is 2,000 km away. In this scenario, our vehicle
traveled along a 4 km curved road with many viaducts. The full
tracking method has 13.1m location accuracy with 5-7 satellites in
sight.

\begin{figure}[t]%[tb]
    \centering
    {\footnotesize
    %\quad
    \shortstack{
            \includegraphics[width=0.6\linewidth]{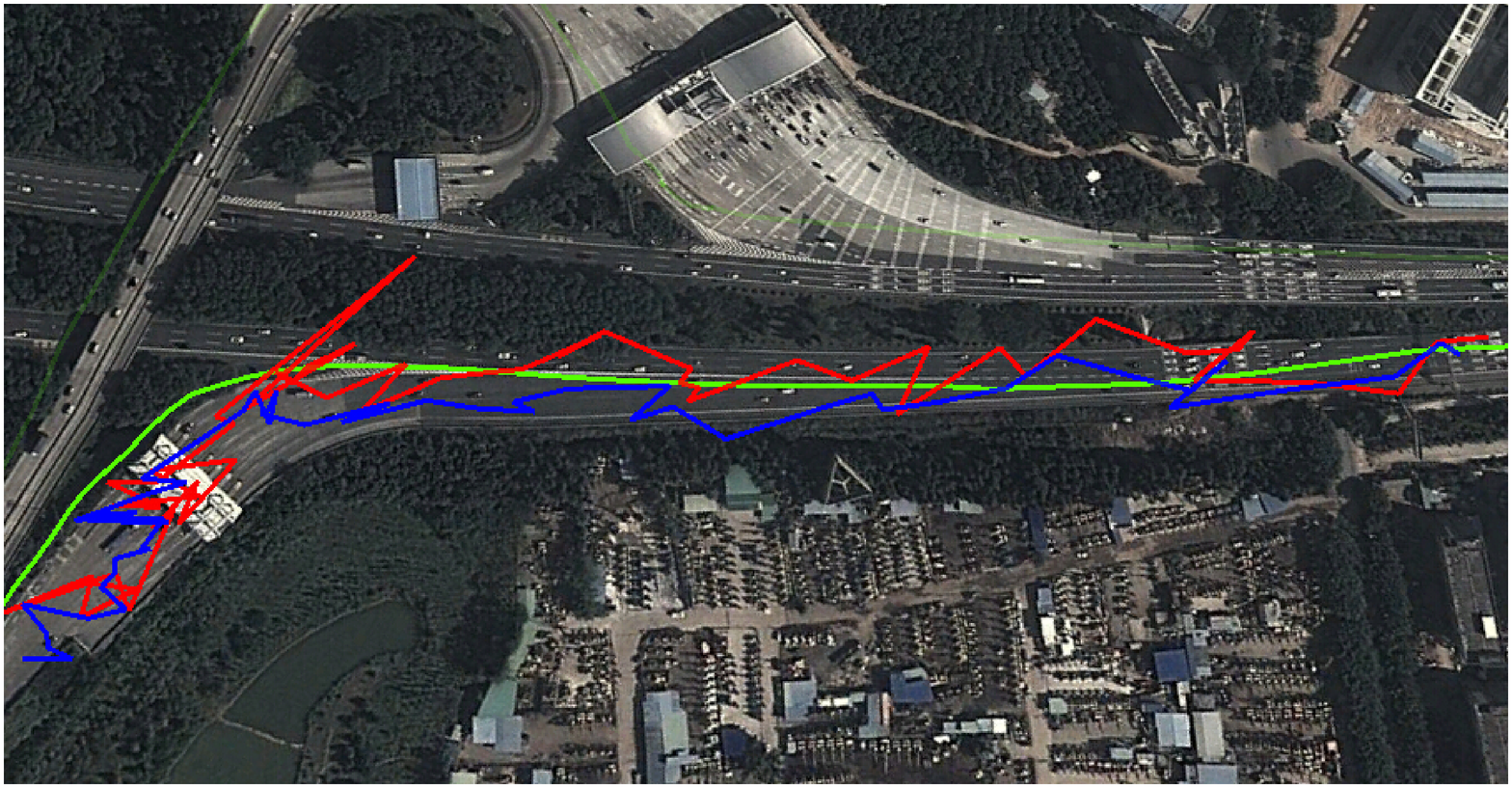}\\
            {(a) Trace one.}
    }
    %\quad
    \shortstack{
            \includegraphics[width=0.6\linewidth]{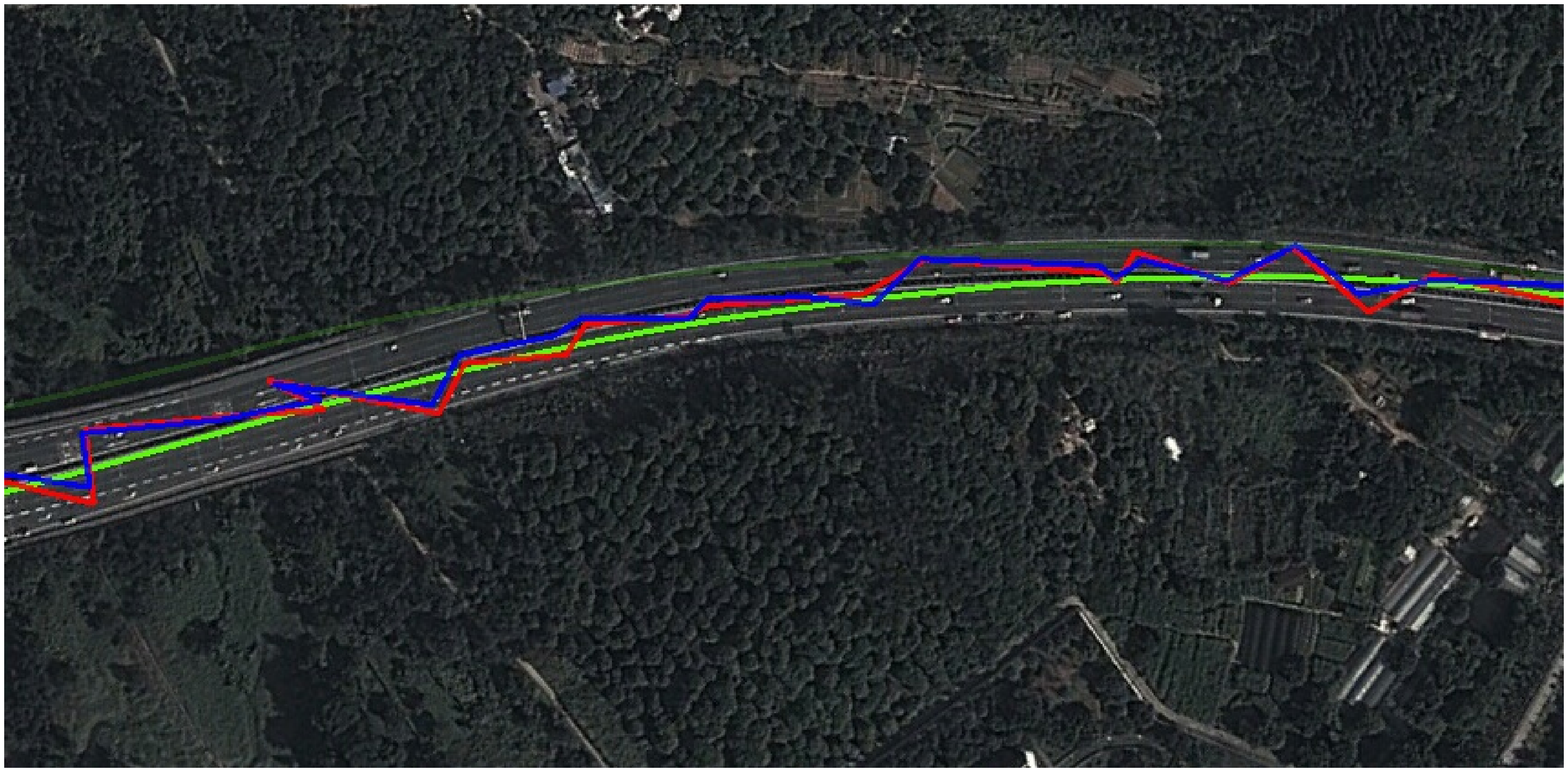}\\
            {(b) Trace two.}
    }
    }
    \caption{The moving trajectories of the GPS receiver under ST (red line),
    FT (blue line), and ground truth (green line).}\vspace{-10pt}
    \label{fig:cmp-gearth}
\end{figure}

\textbf{Evaluation results.} We compare the receiver's position
under selective tracking (ST) and under full tracking (FT) on these
two traces. Figure~\ref{fig:cmp-gearth} presents the calculated
trajectories of the vehicles under ST (red line), FT (blue line), in
comparison with the ground truth (green line). For the first trace, FT produces a mean
position accuracy of 11.9m. ST generates a mean position accuracy
12.7m, with a 23.1\% energy saving.
For the second trace, FT generates 13.1m
location accuracy, while ST shows accuracy of 13.4m, with a 20.9\%
energy saving.

\begin{figure}[t]%[tb]
    \centering
    \includegraphics[width=0.6\linewidth]{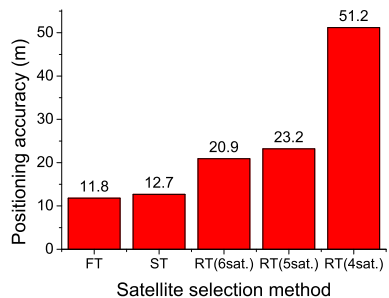}
    \caption{Positioning accuracy of full tracking (FT), selective tracking (ST) and random tracking (RT)
    with different numbers of satellites.}\vspace{-10pt}
    \label{fig:cmp2}
\end{figure}

We also consider a random tracking (RT) method, in which a certain
number of satellites are randomly chosen for tracking.
Figure~\ref{fig:cmp2} demonstrates the positioning accuracy of the
three tracking methods. For the maximum number of randomly chosen
satellites 6, 5, 4, RT provides position accuracy of 20.9m, 23.2m,
and 51.2m, respectively.

\section{Related work}\label{sec:related}
\textbf{Low power location sensing.}
EnLoc~\cite{enloc} computes optimal locations with off-line dynamic
programming, and then selects a localization technology from GPS,
WiFi and GSM for a given energy budge. EnTracked~\cite{entracked}
adjusts the GPS sampling rate based on the estimation and prediction
of system conditions and mobility. RAPS~\cite{rate-adaptive}
presents a rate-adaptive positioning system based on velocity
estimation from historical GPS readings. It also estimates user
movement with a duty-cycled accelerometer, and utilizes Bluetooth
communication with neighboring devices to reduce position
uncertainty. A-Loc~\cite{mobile-loc} builds accuracy models and
energy models for various location sensors, and then designs an
algorithm to determine the most energy efficient sensor for mobile
applications. SensLoc~\cite{sensloc} designs a place detection
algorithm to find contextual information (e.g., home, office) from
sensor signals, and then controls the active duty cycle of a GPS
receiver and other sensors. SmartLoc~\cite{smartloc} is a
localization system to estimate the location and traveling distance
with low power inertial sensors.

In general, these techniques are less power consuming than using GPS
alone, but can be much less accurate in positioning.

\textbf{Computation optimization.} The signal acquisition process
usually consists of a two-dimension Fourier transform, which has a
complexity of $O(n\log n)$ for Fast Fourier Transform (FFT), where
$n$ is the number of signal samples. Hassanieh~\cite{sfft} presents
a sparse Fourier Transform to reduce the complexity from $O(n\log
n)$ to $O(n \sqrt{\log n})$. While the practical improvement is
significant, sparse Fourier transform based method only simplifies
the acquisition progress, and makes very limited contribution to the
whole energy consumption of GPS positioning. Liu et
al.~\cite{cloud-offload,leap} propose to offload the computation
intensive tasks into a cloud server. For each location fix, the GPS
receiver only has to collect and store milliseconds raw GPS signal.
This approach is limited to off-line positioning applications.

\section{Discussion and Conclusion}\label{sec:conclusion}
Although our abstract model in Eq.~\ref{eqn:energyModel} provides an
framework to capture the major software components of a general GPS
receiver, instantiating the model for a specific receiver still
requires the knowledge of the receiver's hardware structure, and
means to measure the power of individual hardware units. This is not
possible for closed and proprietary GPS receivers such as those
found in today's commercial phones. In that case, one can perform
black-box testing to show the impact of certain system parameters
(e.g., update rate) on energy consumption, but the obtainable
information is likely to be very restricted -- for example, it is
not possible to obtain the power breakdown of the different
software/hardware components, which makes it hard to identify the
major energy consumers. As such, our model with Namuru is only the
first step toward a complete understanding of this important module
on mobile devices.

Based on the energy model, we have studied only a simple
optimization to a single system parameter $N$. In the future, we
will consider jointly optimizing multiple procedures and parameters,
and exploiting other location sensors to achieve improved tradeoffs
between positioning accuracy, energy consumption, and solution
applicability.

%%%}
%
\end{document}